\documentclass[aps,pre,amsmath,amssymb,superscriptaddress]{revtex4}

\usepackage{psfig}

\begin{document}

\title{Three-dimensional phase-field simulations of directional solidification}

\author{M. Dejmek}
\affiliation{Laboratoire de Physique de la Mati{\`e}re Condens{\'e}e, \\
CNRS/\'Ecole Polytechnique, 91128 Palaiseau, France}

\author{R. Folch}
\affiliation{Instituut-Lorentz, Universiteit Leiden, 
Postbus 9506, 2300 RA Leiden, The Netherlands}

\author{A. Parisi}
\author{M. Plapp}
\affiliation{Laboratoire de Physique de la Mati{\`e}re Condens{\'e}e, \\
CNRS/\'Ecole Polytechnique, 91128 Palaiseau, France}

\begin{abstract}
The phase-field method has become in recent years the method of
choice for simulating microstructural pattern formation during
solidification. One of its main advantages is that time-dependent
three-dimensional simulations become feasible. This makes it
possible to address long-standing questions of pattern stability. 
Here, we investigate the stability of hexagonal cells and eutectic
lamellae. For cells, it is shown that the geometry of the 
relevant instability modes is determined by the symmetry of
the steady-state pattern, and that the stability limits
strongly depend on the strength of the crystalline anisotropy,
as was previously found in two dimensions. For eutectics,
preliminary investigations of lamella breakup instabilities
are presented. The latter are carried out with a newly developed
phase-field model of two-phase solidification which offers superior
convergence properties.
\end{abstract}

\maketitle

\section{Introduction} 
Microstructural pattern formation during solidification has
been intensely investigated for decades \cite{Kurz},
and yet progress remains to be made on many fascinating questions. 
A central one is pattern selection: For a given material under 
given processing conditions, what is the microstructural
pattern that will emerge, and how is it 
selected out of all other possible structures? 
This is an important issue both from an applied and a fundamental
viewpoint. For the physicist, solidification microstructures
are a classic example of pattern formation out of equilibrium,
and one would like to have a fundamental principle that is capable
to predict the final patterns, as the minimization of an 
appropriate thermodynamic potential does for equilibrium 
situations. However, as of yet, no such general selection 
principle is known for out of equilibrium situations.
While empirical rules for solidification microstructures,
such as the maximum growth speed or the minimum undercooling
criteria, are often very useful in practice, it is often
unclear by what detailed mechanisms the final pattern is 
attained {\em dynamically}. Therefore, it is important to
understand pattern selection more thoroughly starting from the
equations for the {\em evolution} of the solidification front.
\\

\noindent
Recently, modern methods of statistical physics
have allowed theorists to make considerable progress along this
line. In particular, the phase-field method has emerged as a
powerful tool to simulate microstructure evolution. Its main 
advantage over more traditional methods for front
motion is that it avoids an explicit tracking of the interfaces 
with the help of one or several {\em phase fields}, continuous 
auxiliary fields which distinguish between thermodynamic 
phases. This makes it possible to simulate the full dynamics 
of complex morphologies both in two and three dimensions. 
While up to the mid-1990s, this method had remained a
qualitative tool, recent progress has made it possible
to treat {\em quantitatively} some occurrences of
the classic free-boundary problem of solidification
\cite{Karma98,Karma01,Folch03} and to match the results
with theory and experiments \cite{Karma00}.
\\

\noindent
Here, we present three-dimensional phase-field simulations
of the directional solidification of dilute and eutectic 
binary alloys. Both cases are characterized by the existence 
of a continuous family of periodic steady-state solutions 
with different spatial periodicity (spacing): cells for dilute, 
lamellae or rods for eutectic alloys. It has been demonstrated
both in thin-sample experiments \cite{Akamatsu98,Ginibre97}
and two-dimensional simulations \cite{Kopcz96,Karma96,Dresden} 
that no sharp pattern selection occurs: the final spacing depends 
on the growth history, and a range of stable spacings is observed.
Outside of this range of spacings, steady-state solutions generally
exist, but they are {\em unstable}, which explains why they 
cannot be {\em dynamically} selected. Therefore, the limits
of stability determine the observable spacings. The different
instabilities and their onsets have been determined numerically 
for periodic two-dimensional states \cite{Kopcz96,Karma96};
we extend this work to three dimensions, where new instabilities 
can arise.

\section{Dilute Binary Alloy}
Thin-sample directional solidification offers a unique opportunity
to study the influence of crystalline anisotropy on the interface
dynamics experimentally. 
The anisotropy in the sample plane is determined
by the orientation of the three-dimensional crystal with respect
to that plane. If a crystalline (100) axis is normal to the 
sample plane, the in-plane anisotropy is unchanged with respect to 
the three-dimensional situation; in contrast, if the crystal is
oriented with a (111) axis normal to that plane, it is close
to zero. It was observed that shallow cells are
stable only for anisotropic crystals; in the absence of
anisotropy, cell splitting and elimination events occur 
continuously, and the front as a whole never reaches a steady state
\cite{Akamatsu98}.
This is in agreement with a numerical study in two dimensions
by the boundary integral technique \cite{Kopcz96}, 
which indicated that 
the range of stable cellular states strongly depends on anisotropy.
In addition, this study identified the relevant instability modes that
limit the stable range: elimination of every other cell on the
short-spacing side, and a period-doubling oscillatory mode on
the large-spacing side.
\\

\begin{figure}
\centerline{
 \psfig{file=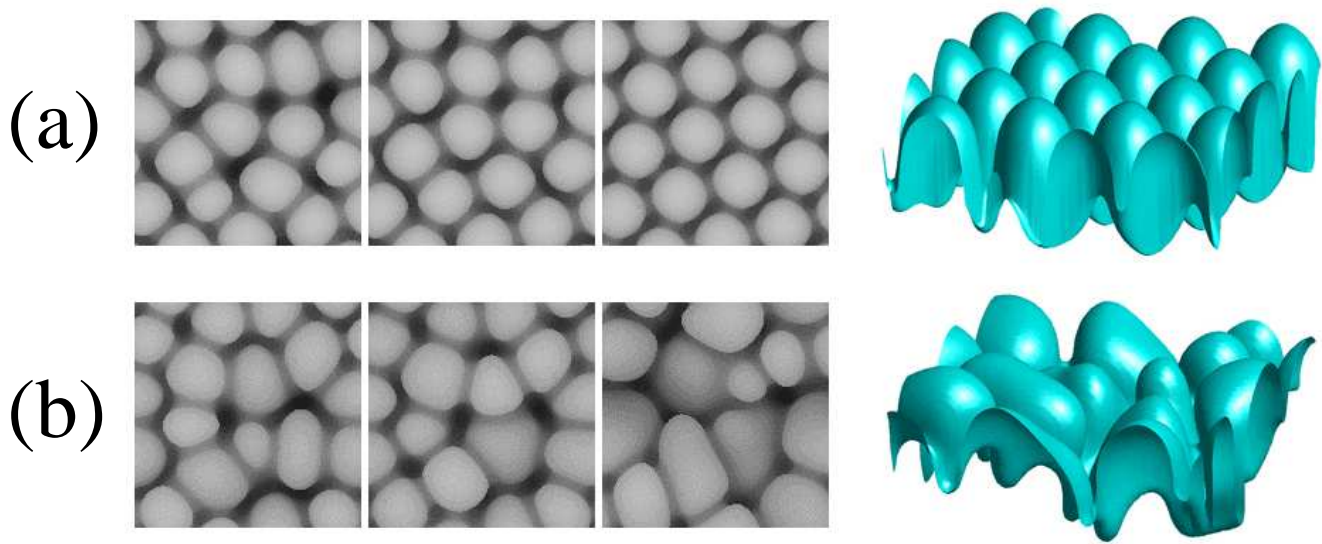,width=\textwidth}}
\caption{Evolution of cellular patterns for different 
anisotropies and otherwise identical parameters: 
$d_0/l_D=2.5\times 10^{-2}$, $\nu=l_T/l_D=12.5$,
where $d_0=\Gamma/(m\Delta c)$ is the chemical
capillary length, $l_D=D/V$ is the diffusion length,
and $l_T=m\Delta c/G$ is the thermal length, with
$\Gamma,m,\Delta c,D,V,G$ being the Gibbs-Thomson
constant, the liquidus slope, the concentration jump
(assumed constant), the impurity diffusion coefficient,
the growth rate, and the temperature gradient, respectively.
(a) $\epsilon_4=0.03$ and (b) $\epsilon_4=0.02$, where $\epsilon_4$
is the strength of the cubic anisotropy. 
Left: top views of the growth front
(the greyscale is proportional to the surface height);
right: 3D view of the interface at the end of the run.}
\label{fig1}
\end{figure}

\noindent
We have extended this investigation to three dimensions using
a phase-field model that has equal diffusivities in both phases, 
and a constant concentration jump (partition coefficient $k=1$).
While this is not very realistic for alloys, the qualitative 
behavior is similar for a one-sided model \cite{Karma01},
and the implementation \cite{PlappCells} is particularly simple.
Simulations that started from a flat steady-state
interface with a small random perturbation are shown in 
Fig.~\ref{fig1}. Clearly, the crystalline anisotropy has
a dramatic effect: whereas the interface spontaneously
organizes into an array of hexagonal cells for sufficiently 
strong anisotropy, the evolution remains unsteady up to the
end of the run for only slightly lower anisotropy.
\\
%
\begin{figure}
\centerline{
 \psfig{file=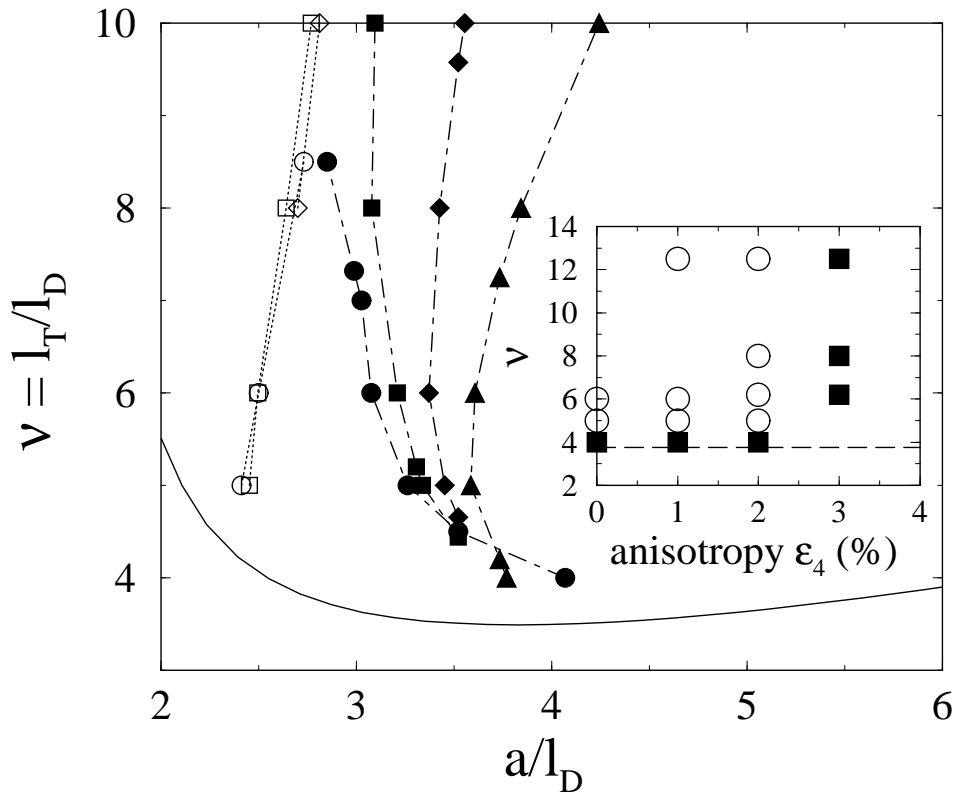,width=.7\textwidth}}
\caption{
Stability boundaries of hexagonal arrays of 
spacing $a$ for different anisotropies: 0\% (circles),
1\% (squares) 2\% (diamonds), and 3\% (triangles).
Arrays are stable between the
limits of cell elimination (dotted lines and open
symbols) and oscillations (dash-dotted lines and
full symbols). The full line is the neutral stability
line of the Mullins-Sekerka instability. Inset: Behavior
for runs as those in Fig. \ref{fig1}. Filled
squares: stable hexagons; open circles: unsteady
evolution; dashed line: Mullins-Sekerka instability
threshold.}
\label{bounds}
\end{figure}

\noindent
To determine the stability boundaries for regular arrays,
we generated states with well-defined spacings by placing
six cells in simulation boxes of appropriate size 
and aspect ratio, and changing control parameters
until an instability occurred. It turns out that the spatial 
structure of the relevant instability modes is determined 
by the symmetry of the underlying steady-state pattern.
A hexagonal lattice can be decomposed into three
equivalent sublattices of $\sqrt{3}$ times larger 
spacing. Instabilities occur when the 
symmetry between the three sublattices is broken.
For small spacings, one out of the three sublattices
is eliminated. For large spacings, the sublattices
start to oscillate, with a phase difference of about
$2\pi/3$ between each other. In both cases, once the
first cell eliminations or splittings occur,
the system follows an unsteady evolution as in 
Fig.~\ref{fig1}b. The stability boundaries are 
plotted in Fig.~\ref{bounds}. The stability 
limit of the oscillatory mode shifts to larger
spacings with increasing anisotropy. This explains the
results of the larger simulations in Fig.~\ref{fig1}: Only when 
a sufficiently large range of stable spacings is available,
the system is able to ``find'' a steady state within a
reasonable simulation time.

\section{Eutectic Alloy}
Coupled eutectic growth leads to lamellar or fibrous 
microstructures. Detailed studies for lamellae in two 
dimensions by thin-sample experiments \cite{Ginibre97} 
and by dynamic boundary-integral simulations \cite{Karma96}
have shown that the range of stable spacings is limited
on the short-spacing side by a long-wavelength lamella
elimination instability, and on the large-spacing side
by oscillatory instabilities of once or twice the
initial spacing. In contrast, very little is known about 
pattern stability in three dimensions, since no reliable
numerical method was available that could handle 
complex composite microstructures.
\\

\noindent
The phase-field method would seem ideally suited to overcome 
this difficulty. However, its use has been hampered until recently 
by the wide range of length scales involved, from an interface 
which is diffuse on a nanometric scale up to a diffusion 
layer of several millimeters for slow solidification. The problem is
that the variation of the phase fields through the diffuse interfaces has
to be resolved by the discretization, which would result in huge
simulation boxes. Fortunately, {\em quantitative} phase-field models
offer the possibility to use an interface thickness $W$ much larger
than that of real solidification fronts, by ensuring that results
are independent of $W$. The only limitation is then that $W$ has
to remain about an order of magnitude smaller than the radius of
curvature of the interface \cite{Karma98,Karma01}.
\\

\begin{figure}
\centerline{
 \psfig{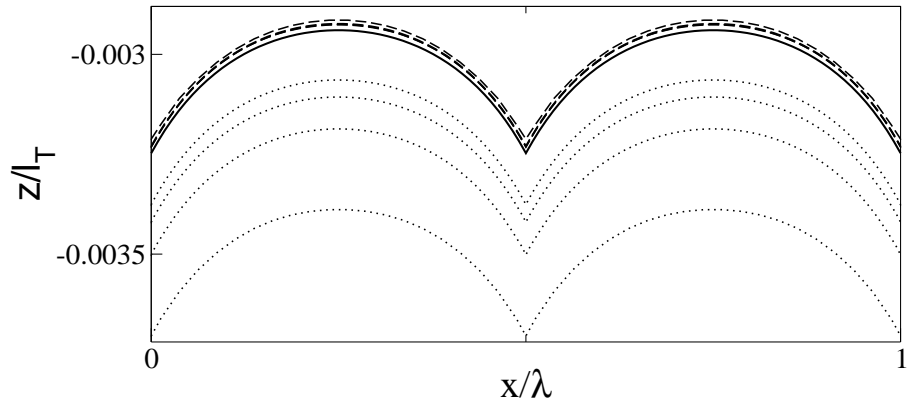}}
\caption{Benchmark simulations of a two-dimensional lamella pair
of spacing $\lambda$ for $d_0/l_D=1.95\times 10^{-5}$, $l_T/l_D=4$, 
$\lambda/\lambda_m=1$, where $\lambda_m$ is the minimum-undercooling
spacing. Solid line: boundary integral. Dotted lines: phase-field
model of Ref. \cite{Dresden}. Dashed lines: present phase-field model. 
Four curves at $\lambda/W=32$, $64$, $96$ and $128$
shown per model; the results depend on $W$ for the generic model,
but superimpose for $\lambda/W\ge 64$ for our new model.}
\label{figconv}
\end{figure}

\noindent
We have recently developed a quantitative
model for two-phase solidification with vanishing 
solute diffusion in the two solid phases \cite{Folch03}.
As a test, we have performed simulations in two 
dimensions at the eutectic composition of a model 
alloy that has a completely symmetric phase 
diagram (generalization to other compositions and 
arbitrary phase diagrams is straightforward \cite{Folch03}),
and compared the results to a boundary integral
calculation with the code of Ref.~\cite{Karma96}.
As can be seen in Fig.~\ref{figconv}, our model
produces precise results that are independent 
of the interface thickness for $\lambda/W\ge 64$,
where $\lambda$ is the lamellar spacing,
whereas those of a generic model make appreciable 
errors and depend on $W$.
\\

\begin{figure}
\centerline{
 \psfig{file=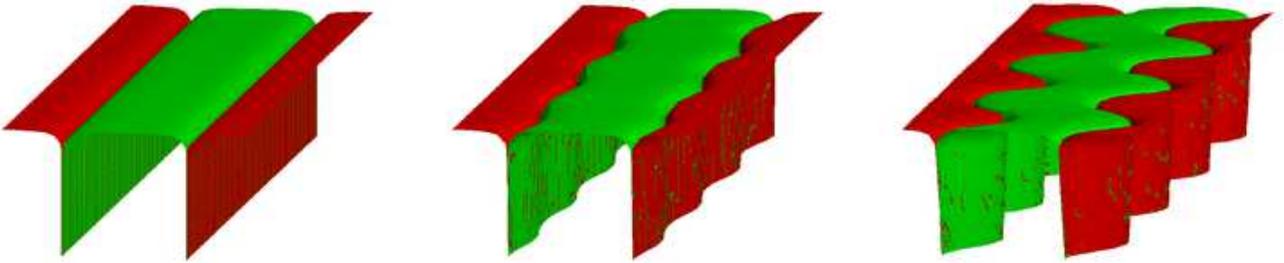,width=\textwidth}}
\caption{Snapshot pictures of a three-dimensional lamella
that undergoes breakup instabilities. Parameters: $d_0/l_D=10^{-3}$,
$l_T/l_D=4$, $\lambda/\lambda_m=1.6$.}
\label{fig3d}
\end{figure}

\begin{figure}
\centerline{
 \psfig{file=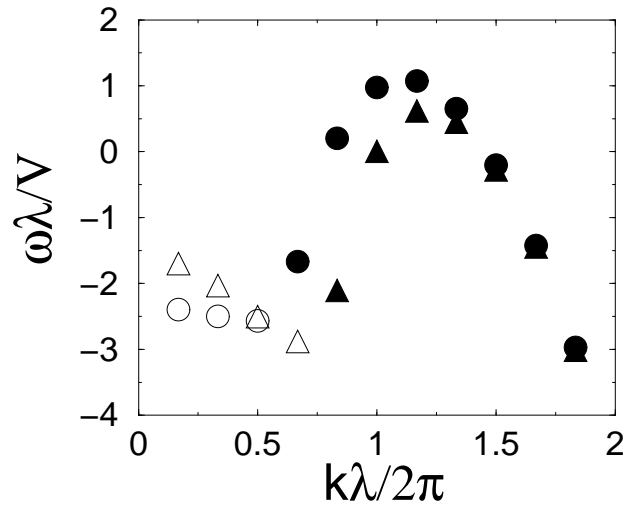,width=.45\textwidth}}
\caption{Stability spectrum (growth rate $\omega$ versus wave
number $k$) of the run shown above. Circles: modulations 
of the mid-line position; triangles: modulations of the 
lamella thickness. Open symbols: oscillatory modes; filled
symbols: exponential modes.}
\label{figspect}
\end{figure}

\noindent
In Fig.~\ref{fig3d}, we show a three-dimensional simulation
of an unstable lamella with reflecting boundary conditions on 
all sides. The lateral dimensions are $56\times 168$
grid points, and the diffusion field along the growth direction 
is calculated using a multi--grid method with coarser 
and coarser grids away from the interface. Undulations 
develop with time, and finally the lamella breaks up,
leaving behind a maze of the two phases (not shown).
This evolution can be further analyzed to characterize the
instability. First, the positions of the triple lines are
extracted. Then, we calculate the position of the mid-line
and the local width of the lamella. These two variables
correspond to two different instability modes: the former
to an undulation of the whole lamella, the latter to a
pinching mechanism. After a Fourier transform, it can
be checked that modes of different wave number $k$ along
the lamella grow exponentially with a $k$-dependent growth
rate $\omega(k)$. The extracted stability spectrum is plotted
in Fig.~\ref{figspect}. The ``undulating'' instability grows 
faster, and indeed it is this mode that shows up in Fig.~\ref{fig3d}.
It should be noted that this three-dimensional instability
occurs for parameters where the two-dimensional system is
{\em stable} both in simulations and experiments. Therefore,
it can already be concluded from this preliminary simulation
that the range of stable spacings is considerably smaller
in three dimensions than in two.

\section{Conclusion}
Phase-field modeling offers a unique opportunity to investigate 
pattern stability and pattern selection in three dimensions.
Here, we have focused on results for generic phase diagrams, 
but the method can be carried over to the study of particular
materials; simulations for the alloy of Refs.~\cite{Akamatsu98,Ginibre97} 
are currently in progress. This will enable us to carry out
critical comparisons of simulations, experiments, and theory,
which constitutes a particularly promising way to elucidate 
many long-standing questions concerning microstructural 
pattern formation during solidification.

\section{Acknowledgments}
We thank Alain Karma for many useful discussions, and 
for the boundary integral code. This work was supported
by the Centre National d'Etudes Spatiales, France.

\end{document}